\documentclass[12pt,preprint]{aastex}
\def\ltsima{$\; \buildrel < \over \sim \;$}
\def\gtsima{$\; \buildrel > \over \sim \;$}
\def\lsim{\lower.5ex\hbox{\ltsima}}
\def\gsim{\lower.5ex\hbox{\gtsima}}
\def\lapp{\ifmmode\stackrel{<}{_{\sim}}\else$\stackrel{<}{_{\sim}}$\fi}
\def\gapp{\ifmmode\stackrel{>}{_{\sim}}\else$\stackrel{<}{_{\sim}}$\fi}

\usepackage{amsmath}
\usepackage{textcomp}
\usepackage{amssymb}

\newdimen\minuswidth    %define @ width of minus sign for tables
\setbox0=\hbox{$-$}
\minuswidth=\wd0
\catcode`@=\active
\def@{\kern\minuswidth}
\setbox0=\hbox{\rm0}
\shorttitle{} 
\shortauthors{Massari et al.}
 
\begin{document} 
\title{Chemical and kinematical properties of Galactic bulge stars surrounding the 
stellar system Terzan~5\footnote{ 
Based on FLAMES observations collected at the European Southern Observatory, 
proposal numbers 087.D-0716(B), 087.D-0748(A) and 283.D-5027(A) and at the W. M. Keck Observatory, which 
is operated as a scientific partnership among the California Institute of 
Technology, the University of California, and the National Aeronautics and Space Administration.
The Observatory was made possible by the generous financial support of the W. M. Keck Foundation.}
}

\author{
D. Massari\altaffilmark{2},
A. Mucciarelli\altaffilmark{2},
F. R. Ferraro\altaffilmark{2},
L. Origlia\altaffilmark{3},
R. M. Rich\altaffilmark{4},
B. Lanzoni\altaffilmark{2},
E. Dalessandro\altaffilmark{2},
R. Ibata\altaffilmark{5},
L. Lovisi\altaffilmark{2},
M. Bellazzini\altaffilmark{3},
D. Reitzel\altaffilmark{4}
}
\affil{\altaffilmark{2} Dipartimento di Fisica e Astronomia, Universit\`a degli Studi
di Bologna, v.le Berti Pichat 6/2, I$-$40127 Bologna, Italy}
\affil{\altaffilmark{3}INAF-Osservatorio Astronomico di Bologna, via
  Ranzani 1, 40127, Bologna, Italy} 
\affil{\altaffilmark{4} 
Department of Physics and Astronomy, Math-Sciences 8979, UCLA, Los Angeles, CA 90095-1562, USA} 
\affil{\altaffilmark{5} 
Observatoire Astronomique, Universit\'e de Strasbourg, CNRS, 11, rue de l'Universit\'e. 
F-67000 Strasbourg, France}

\date{29 Apr, 2014}

\begin{abstract}
As part of a study aimed at determining the kinematical and chemical
properties of Terzan 5, we present the first characterization of the
bulge stars surrounding this puzzling stellar system.  We observed 615
targets located well beyond the tidal radius of Terzan 5 and we found
that their radial velocity distribution is well described by a
Gaussian function peaked at $<v_{{\rm rad}}>=+21.0\pm4.6$
km$\,$s$^{-1}$ and with dispersion $\sigma_{v}=113.0\pm2.7$
km$\,$s$^{-1}$.  This is the one of the few high-precision spectroscopic survey
of radial velocities for a large sample of bulge stars in such a low
and positive latitude environment (b=+1.7\textdegree).  We found no
evidence for the peak at $<v_{{\rm rad}}>\sim+200$ km$\,$s$^{-1}$ found in
\cite{nidever}. 
The strong contamination of many observed spectra by TiO bands 
prevented us from deriving the iron abundance for the entire spectroscopic sample, 
introducing a selection bias. The metallicity distribution was finally derived for a sub-sample of 112 stars
in a magnitude range where the effect of the selection bias is negligible. The distribution
is quite broad and roughly peaked at solar metallicity
([Fe/H]$\simeq+0.05$ dex) with a similar number of
stars in the super-solar and in the sub-solar ranges.  
The population number ratios in different metallicity ranges agree well with those observed
in other low-latitude bulge fields suggesting ($i$) the possible presence of a plateau
for $|$b$|<4$\textdegree~ for the ratio between stars in the super-solar ($0<$[Fe/H]$<0.5$ dex)
and sub-solar ($-0.5<$[Fe/H]$<0$ dex) metallicity ranges; ($ii$) a severe drop of the
metal-poor component ([Fe/H]$<-0.5$) as a function of Galactic latitude.
\end{abstract}
 
\keywords{galaxy: bulge;\ stars:\ abundances;\ stars:\ radial velocities;\ techniques:\ spectroscopic}

\section{INTRODUCTION}
Terzan 5 is a stellar system located in the bulge of our Galaxy,
at a distance of 5.9 kpc (\citealt{valenti07, valenti10}). Because of
the large and spatially varying extinction (\citealt{massari})
critically affecting any optical observation of the system, its true
nature remained hidden until near infrared (NIR) observations revealed
its peculiar properties. In fact, by using J and K band data acquired
with the Multi-conjugate Adaptive optics Demonstrator (MAD) mounted at
the Very Large Telescope, \cite{f09} discovered the presence of two
distinct stellar populations.  The analysis of high-resolution IR
spectra (\citealt{origlia97}) obtained with NIRSPEC at the Keck II
telescope, demonstrated that the two populations have different
metallicities, the metal-poor component being sub-solar with
[Fe/H]$=-0.25$ dex and the metal-rich one having [Fe/H]$=+0.27$ dex
(\citealt{f09,origlia}).  Recently, we discovered even a third,
more metal-poor population with [Fe/H]$\simeq-0.8$ dex (\citealt{o13};
Davide Massari et al. in preparation).  The analysis of the
$\alpha$-elements also revealed that the two metal-poor populations
are $\alpha$-enhanced, and therefore formed from gas mainly enriched
by type II supernovae (SNII).  Instead the metal-rich component has
a solar [$\alpha$/Fe] ratio and thus we infer that it formed from 
gas further polluted by type Ia supernovae (SNIa).  
This abundance pattern, with the
$\alpha$-elements being enhanced up to solar metallicity and then
progressively decreasing towards solar values \citep[see][]{origlia},
is strikingly similar to what is typically observed for the bulge
field stars (see for instance \citealt{mcw94, ful07, hill11, rich12,
  johnson13, ness13}). However it must be noticed that the precise
metallicity at which the knee in the [$\alpha$/Fe] {\it vs} [Fe/H]
trend occurs is still controversial, since measurements are quite
scattered, and different elements as well as different studies provide
somewhat different answers.

In order to investigate the kinematical and chemical properties of
Terzan 5, we have collected spectra for more than 1600 stars in its
direction. In this paper we focus on the properties of the bulge field
population surrounding the system, with the aim of providing crucial
information for future studies of both Terzan 5 and the bulge
itself. In fact, this is a statistically significant sample of field
stars which can be used for the purpose of decontaminating Terzan 5
population from non-members.  Moreover, it is one of the few large
samples of bulge stars spectroscopically investigated at low and
positive latitudes (b$=+1.7$\textdegree), thus allowing interesting
comparisons with other well studied bulge regions.

The paper is organized as follows. In Section \ref{obs} we present the
spectroscopic sample analyzed in this work.  In Section \ref{vrad} we
describe the analysis and the results concerning the radial
velocities.  In Section \ref{iron} we describe the chemical analysis
and the metallicity distribution of the sub-sample of stars for which
we determined iron abundances, and in Section \ref{summary} we
summarize the results obtained.

\section{THE SAMPLE}
\label{obs}
\begin{figure}%[!htb]
\plotone{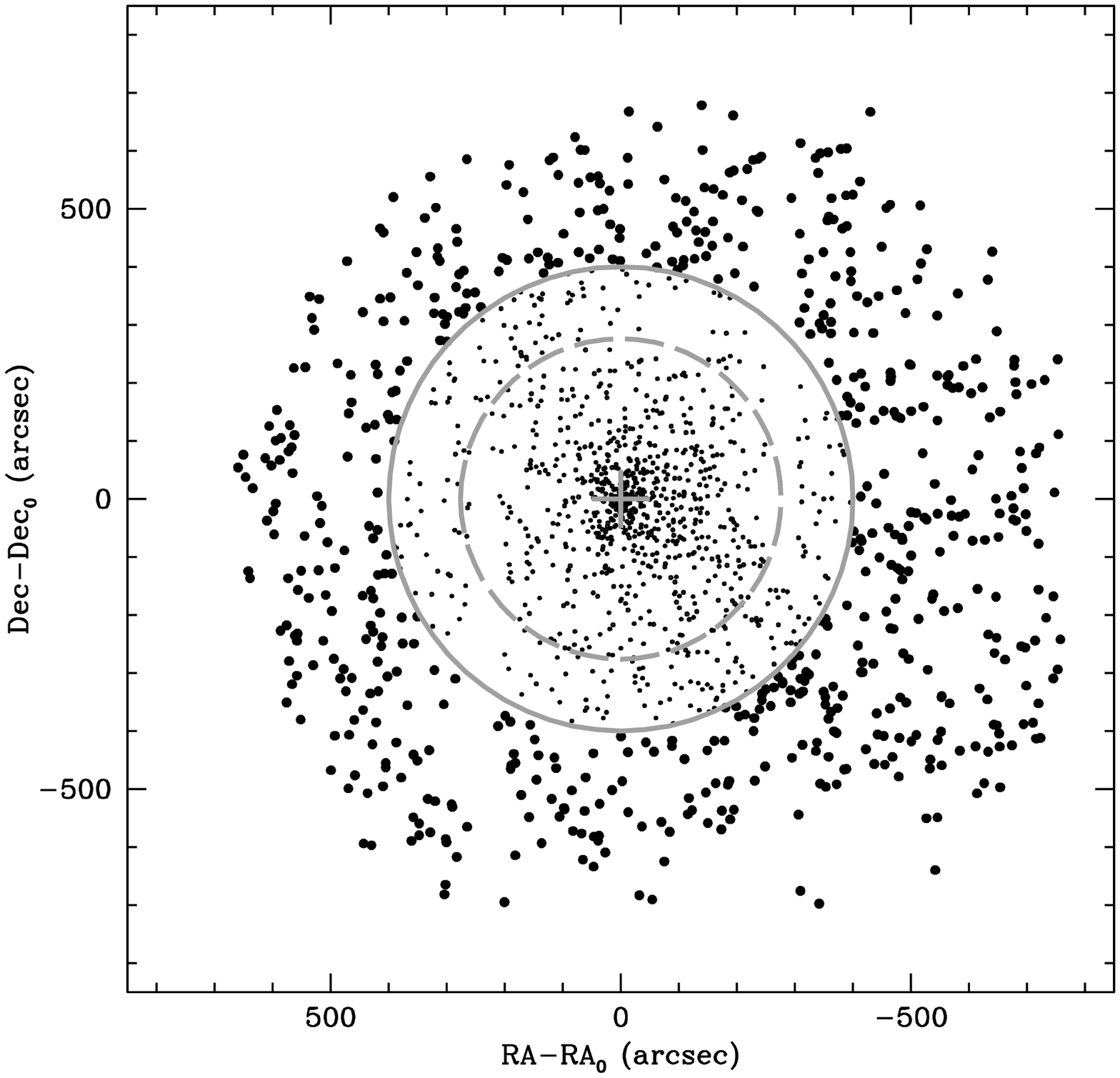}
\caption{\small Spatial distribution of all the targets in our
  spectroscopic survey in the direction of Terzan 5. The center of
  gravity and tidal radius ($r_t\simeq 300\arcsec$) of Terzan 5 (from
  \citealt{l10}) are marked with a gray cross and a gray dashed
  circle, respectively. The targets discussed in the paper (shown as
  filled black circles) are all located at more than 400\arcsec ~from
  the center (gray solid circle), well beyond the Terzan 5 tidal
  radius.}
\label{map}
\end{figure}

This study is based on a sample of 1608 stars within a radius of 800\arcsec
from the center of Terzan 5 ($\alpha_{{\rm J}2000}=17^{\rm
  h}:48^{\rm m}:4^{\rm s\!.}85$, $\delta_{{\rm
    J}2000}=-24^{\circ}:46\arcmin:44\farcs6$; see \citealt{f09, l10}).
While the overall
survey will be described in a forthcoming paper (Francesco
Ferraro et al. in preparation), in this work we focus on a sub-sample
of stars representative of the field population surrounding Terzan
5. Given the value of the tidal radius of the system
(r$_{t}\simeq300\arcsec$; \citealt{l10, miocchi13}), we conservatively
selected as genuine field population members all the targets more
distant than $400\arcsec$ from the center of Terzan 5.  This
sub-sample is composed of $615$ stars belonging to two different
datasets obtained with FLAMES \citep{pasquini} at the ESO Very Large
Telescope (VLT) and with DEIMOS (\citealt{faber}) at the Keck II
Telescope.  Each target has been selected from the ESO-WFI optical
catalog described in \cite{l10}, along the brightest portion of the
red giant branch (RGB), with magnitudes brighter than $I<18.5$.  In
order to avoid contamination from other sources, in the selection
process of the spectroscopic targets we avoided stars with bright
neighbors (I$_{neighbor}<{\rm I_{star}}+1.0$) within a distance of
2\arcsec.  The spatial distribution of the entire sample is shown in
Figure \ref{map}, where the selected field members are shown as large
filled circles.

The FLAMES dataset was collected under three different programs:
087.D-0716(B), PI: Ferraro; 087.D-0748(A), PI: Lovisi; and
283.D-5027(A), PI: Ferraro.  All the spectra were acquired in the
GIRAFFE/MEDUSA mode, allowing the allocation of 132 fibers across a
$25\arcmin$ diameter field of view (FoV) in a single pointing. We used
the GIRAFFE setup HR21, with a resolving power of 16200 and a spectral
coverage ranging from 8484 \AA{} to 9001 \AA{}.  This grating was
chosen because it includes the prominent Ca~II triplet lines, which
are excellent features to measure radial velocities also in relatively
low signal-to-noise ratio (SNR) spectra. Other metal lines (mainly
Fe~I) lie in this spectral range, thus allowing a direct measurement
of the iron abundance.  Multiple exposures (with integration times
ranging from 1500 to 3000 s, according to the magnitude of the
targets) were secured for the majority of the stars, in order to reach
SNR$\sim$30 even for the faintest ($I\sim 18.5$) targets.  The data
reduction was performed with the FLAMES-GIRAFFE
pipeline\footnote{http://www.eso.org/sci/software/pipelines/},
including bias-subtraction, flat-field correction, wavelength
calibration with a standard Th-Ar lamp, re-sampling at a constant
pixel-size and extraction of one-dimensional spectra.  Since a correct
sky subtraction is particularly crucial in this spectral range
(because of the large number of O$_{2}$ and OH emission lines), 15-20
fibers were used to measure the sky in each exposure. Then a master
sky spectrum was obtained from the median of these spectra and it was
subtracted from the target spectra.  Finally, all the spectra were
shifted to zero-velocity and in the case of multiple exposures they
were co-added.

The DEIMOS dataset was acquired using the 1200 line/mm grating coupled
with the GG495 and GG550 order-blocking filters, covering the
$\sim$6500-9500 \AA{} spectral range at a resolution of R$\sim$7000 at
$\sim$8500 \AA{}.  The DEIMOS FoV is $16$\arcmin$\times5$\arcmin,
allowing the allocation of more than 100 slits in a single mask.  The
observations were performed with an exposure time of 600 s, securing
SNR$\sim$50-60 spectra for the brightest targets and achieving
SNR$\sim$15-20 for the faintest ones (I$\sim17$).  The spectra have
been reduced by means of the package developed for an optimal
reduction and extraction of DEIMOS spectra and described in
\citet{ibata11}.

\section{RADIAL VELOCITIES}
\label{vrad}
\begin{figure}%[!htb]
\plotone{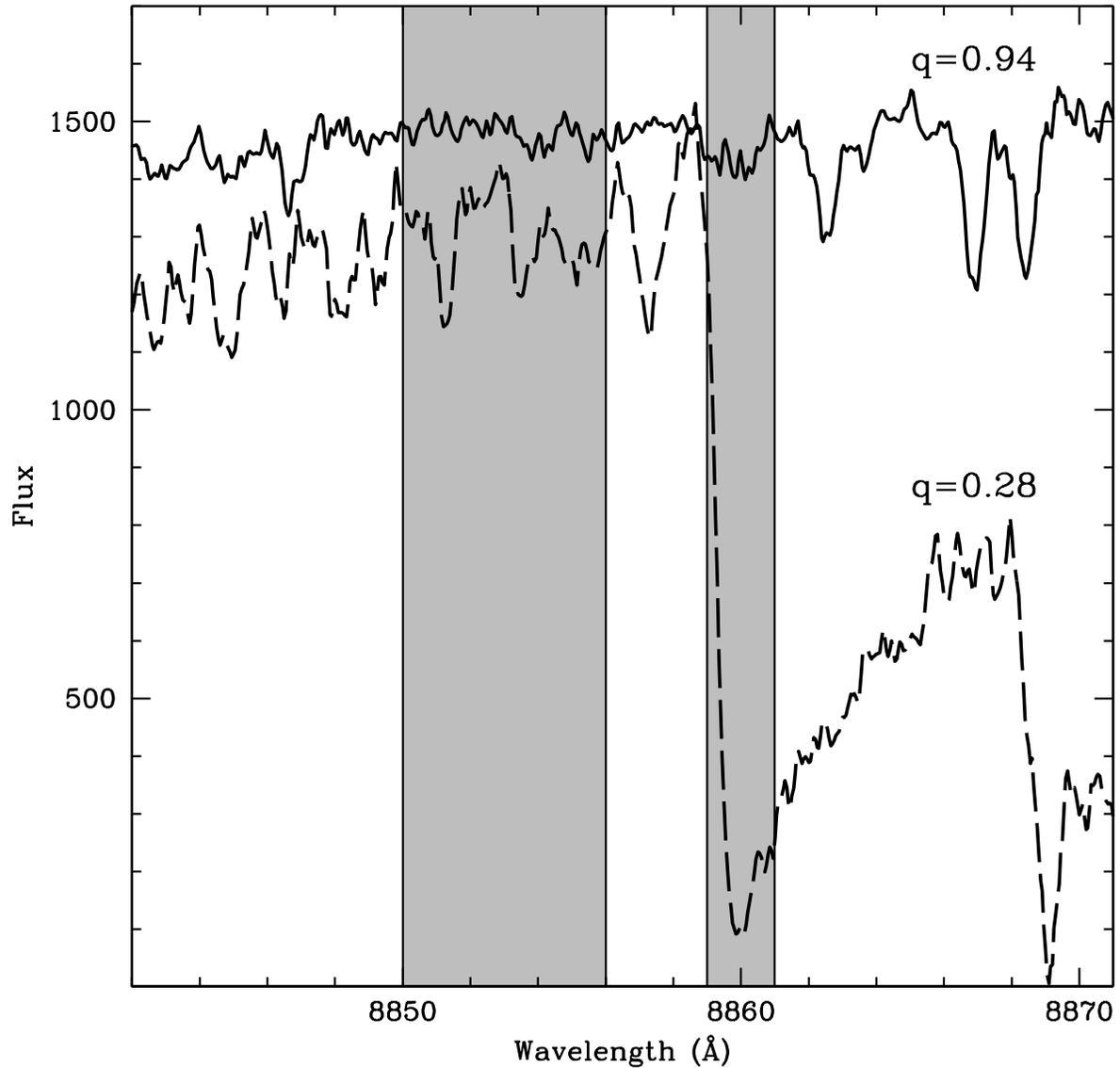}
\caption{\small Examples of spectra poorly (solid line) and
  severely (dashed line) affected by the TiO molecular bands
  ($\lambda>8860$ \AA).  The gray regions highlight the wavelength
  ranges adopted to compute the $q$-parameter defined in the text
  (Section \ref{iron}).  For these two spectra very different values
  of $q$ have been obtained: $q=0.94$ and $q=0.28$ for the solid and
  dashed spectrum, respectively. According to the adopted selection
  criterion (q$>0.6$), the iron abundance has not been computed from
  the latter.  }
\label{tiosel}
\end{figure}

Radial velocities ($v_{\rm rad}$) for the target stars were measured
by cross-correlating the observed spectra with a template of known
velocity, following the procedure described in \citet{tonry} and
implemented in the FXCOR software under IRAF.  As templates we adopted
synthetic spectra computed with the SYNTHE code \citep{sbordone04}.
For most of the stars the cross-correlation procedure is performed in
the spectral region $\sim$8490-8700 \AA{}, including the prominent
Ca~II triplet lines that can be well detected also in noisy spectra.
For some very cool stars, strong TiO molecular
bands dominate this spectral region, preventing any reliable
measurement of the Ca features (Figure \ref{tiosel} shows the
comparison between two FLAMES spectra, with and without strong
molecular bands). In these cases, the radial velocity was measured
from the TiO lines by considering only the spectral region around the
TiO bandhead at $\sim$8860 \AA{} and by using as template a synthetic
spectrum including all these features.  Because several stars show
both the Ca~II triplet lines and weak TiO bandheads, for some of them
the radial velocity has been measured independently using the two
spectral regions. We always found an excellent agreement between the
two measurements, thus ruling out possible offsets between the two
$v_{\rm rad}$ diagnostics.

For the FLAMES dataset, where multiple exposures were secured for most
of the stars, radial velocities were obtained from each exposure
independently. The final radial velocity is computed as the weighted
mean of the individual velocities (each corrected for its own
heliocentric velocity), by using the formal errors provided by FXCOR
as weights.  For the DEIMOS spectra, we checked for possible velocity
offsets due to the mis-centering of the target within the slit
\citep[see the discussion about this effect in][]{simon}, through the
cross-correlation of the A telluric band (7600-7630 \AA{}). We found
these offsets to be of the order of few km$\,$s$^{-1}$.  The
uncertainty on the determination of this correction (always smaller
than 1 km$\,$s$^{-1}$) has been added in quadrature to that provided
by FXCOR.  The typical final error on our measured v$_{{\rm rad}}$ is
$\sim1.0$ km$\,$s$^{-1}$.

\begin{figure}%[!htb]
\plotone{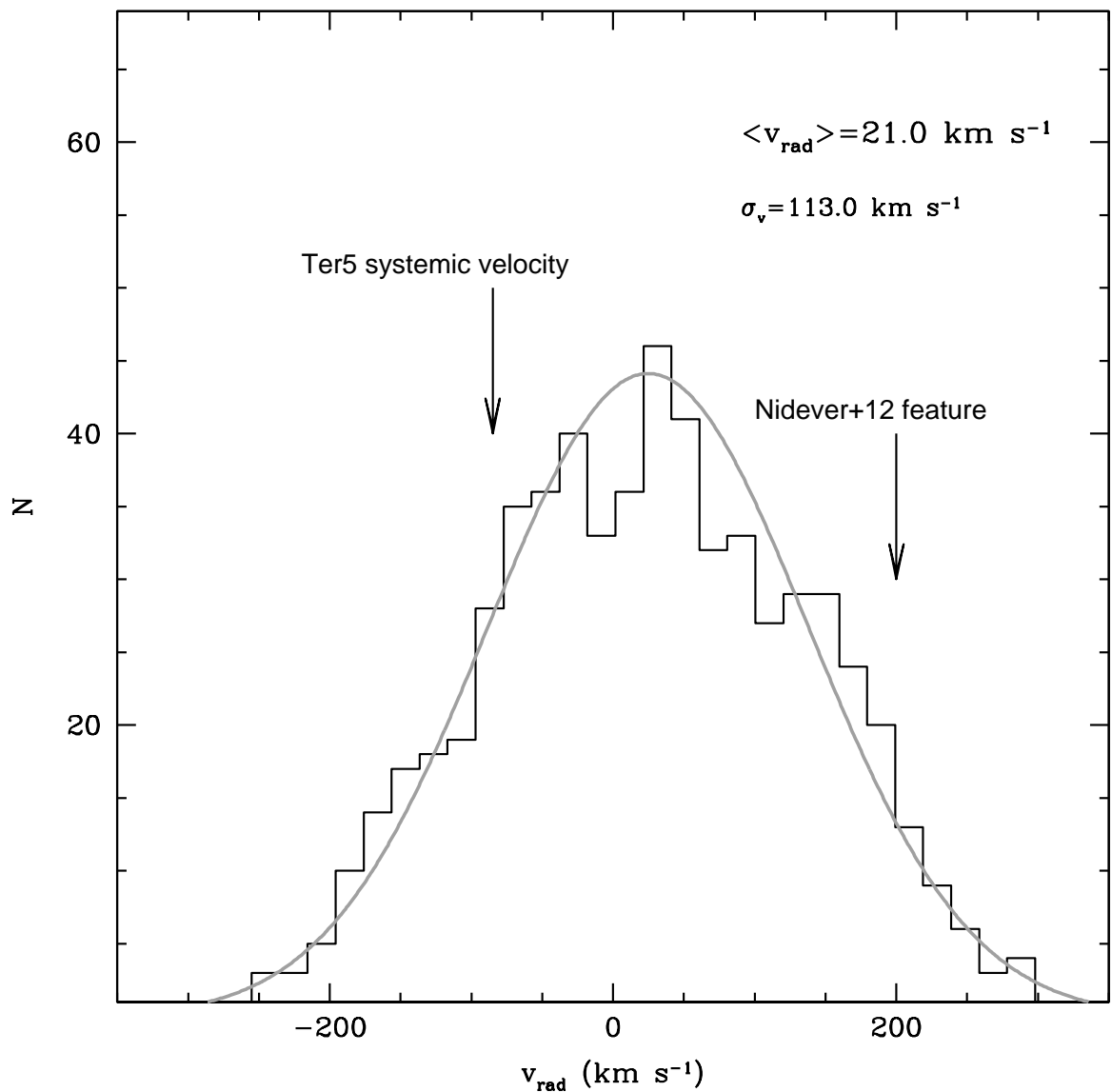}
\caption{\small Radial velocity distribution for the 615 spectroscopic
  targets at $r>400\arcsec$. The mean value and dispersion are
  indicated. The bin size (20 km$\,$s$^{-1}$) is the same as that
  adopted by \cite{nidever}, but in our case no high velocity
  sub-components are found. The systemic velocity of Terzan 5 
(v$_{{\rm rad}}\simeq-83$ km$\,$s$^{-1}$) and the location of the subcomponent
found in \cite{nidever} (v$_{{\rm rad}}\simeq200$ km$\,$s$^{-1}$) are also marked 
with the black arrows for sake of comparison.}
\label{vraddist}
\end{figure}

The distribution of the measured $v_{\rm rad}$ for the 615 targets is
shown in   Figure \ref{vraddist}.  It ranges from
$-264.0$ km s$^{-1}$ to $+303.9$ km s$^{-1}$.  By using a
Maximum-Likelihood procedure, we find that the Gaussian function that
best describes the distribution has mean $\langle v_{\rm
  rad}\rangle=21.0\pm4.6$ km s$^{-1}$ and $\sigma_{v}=113.0\pm2.7$ km
s$^{-1}$.
We converted radial velocities to Galactocentric velocities ($v_{\rm
  GC}$) by correcting for the Solar reflex motion (220 km s$^{-1}$;
\citealt{kerrlyn86}) and assuming as peculiar velocity of the Sun in
the direction (l,b)$=$(53\textdegree,25\textdegree) $v=18.0$ km
s$^{-1}$ \citep{schonrich}. The conversion equation is then:
\begin{equation}
 v_{{\rm GC}}=v_{{\rm rad}}+220[\sin(l)\cos(b)]+18[\sin(b)\sin(25)+\cos(l)\cos(25)\cos(l-53)],
\end{equation}
where velocities are in km s$^{-1}$, and (l,b)=(3.8\textdegree,
1.7\textdegree) is the location of Terzan 5.
The Galactocentric velocity distribution estimated in this way peaks at $<v_{{\rm GC}}>=47.7\pm4.6$ km$\,$s$^{-1}$.

This value turns out to be in good agreement with the values found in the context of three recent kinematic surveys of the Galactic
bulge: the BRAVA survey (\citealt{rich07}), the ARGOS survey (\citealt{freeman13}) and the GIBS survey (\citealt{zoccali14}).
In fact, in fields located close to the Galactic Plane and with Galactic longitude 
as similar as possible to ours, \citet[][see also Kunder et al. 2012]{howard08} found $<v_{{\rm GC}}>=53.0\pm10.3$ km$\,$s$^{-1}$ at 
(l,b)=(4\textdegree,-3.5\textdegree) for BRAVA, \cite{ness13kin} found $<v_{{\rm GC}}>=44.4\pm3.8$ km$\,$s$^{-1}$ 
at (l,b)=(5\textdegree,-5\textdegree) for ARGOS and \cite{zoccali14} obtained $<v_{{\rm GC}}>=55.9\pm3.9$ km$\,$s$^{-1}$ 
at (l,b)=(3\textdegree,-2\textdegree) for GIBS.
Thus, all the measurements agree within the errors with our result.
As for the velocity dispersion, our estimate ($\sigma_{v}=113\pm2.7$ km$\,$s$^{-1}$) agrees well with the result of 
\cite{kunder12}, who found $\sigma_{v}=106$ km$\,$s$^{-1}$, and with that of \cite{zoccali08} who measured
$\sigma_{v}=112.5\pm6.4$ km$\,$s$^{-1}$. Instead, it is larger than that quoted 
in \cite{ness13kin} $\sigma_{v}=92.2\pm2.7$ km$\,$s$^{-1}$. Since their field is the farthest from ours among
those selected for the comparison, we ascribe such a difference to the different location in the bulge.
As a further check, we used the Besan\c{c}on Galactic model
(\citealt{robin}) to simulate a field with the same size of our
photometric sample (i.e., the WFI FoV) around the location of Terzan
5, and we selected all the bulge stars lying within the same color and
magnitude limits of our sample. The velocity dispersion for these
simulated stars is $\sigma_{v}=119$ km s$^{-1}$, in agreement with
our estimate.

It is worth mentioning that \cite{nidever} identified a high velocity
($v_{{\rm rad}}\sim$200 km s$^{-1}$) sub-component that accounts for about
10\% of their entire sample of $\sim4700$ bulge stars.  Such a feature
has been found in eight fields located at
-4.3\textdegree$<$b$<2$\textdegree and
4\textdegree$<$l$<14$\textdegree.  \cite{nidever} suggest that such a
high-velocity feature may correspond to stars in the Galactic bar
which have been missed by other surveys because of the low latitude of
the sampled fields.  However, as is evident in Figure \ref{vraddist}
(where we adopted the same bin-size used in Fig. 2 of \cite{nidever}
for sake of comparison), we do not find neither high-velocity peak
nor isolated substructures in our sample, despite its low latitude.  
In fact, the skewness calculated for our distribution is $-0.02$, 
clearly demonstrating its symmetry.  Also \cite{zoccali14} did not find 
any significant peak at such large velocity in the recent GIBS survey.

\section{METALLICITIES}\label{iron}

For a sub-sample of $284$ stars we were able to also derive
metallicity. As already pointed out in Section \ref{vrad},
spectra of cool giants are affected by the presence of prominent TiO molecular
bands. These bands make particularly uncertain the determination of
the continuum level.  While this effect has no consequences on the
determination of radial velocities, it could critically affects the
metallicity estimate. Therefore we limited the metallicity analysis
only to stars whose spectra suffer from little contamination from the
TiO bands. In order to properly evaluate the impact of the TiO bands
in the considered wavelength range, we performed a detailed analysis
of a large set of synthetic spectra and we defined a parameter $q$ as
the ratio between the flux of the deepest feature of the TiO bandhead
at 8860 \AA{} (computed as the minimum value in the spectral range
8859.5 \AA{}$<\lambda<$8861 \AA{}) and the continuum level measured
with an iterative 3 $\sigma$-clipping procedure in the adjacent
spectral range, 8850\AA{}$<\lambda<$8856 \AA{} (see the shaded regions
of Figure \ref{tiosel}).  

We found that the continuum level of synthetic spectra for 
stars with $q>0.8$ is slightly ($<2\%$) affected by TiO bands over 
the entire spectral range, while for stars with $0.6<q<0.8$ the region
marginally ($<5\%$) affected by the contamination is confined between 8680
\AA{} and 8850 \AA. Instead, stars with $q<0.6$ have no useful spectral
ranges (where at least one of the Fe~I in our linelist falls) 
with TiO contamination weak enough to allow a 
reliable chemical analysis. We therefore analyzed
targets with $q>0.8$ (counting 126 objects) using the full linelist
(see Section \ref{analysis}), while for targets with $0.6<q<0.8$ (158
objects) only a sub-set of atomic lines lying in the safe spectral
range $8680-8850$ \AA{} has been adopted.  All targets with $q<0.6$
(329 stars) have been excluded from the metallicity analysis.  Hence
the metallicity analysis is limited to 284 stars (corresponding to
$\sim 46\%$ of the entire sample observed in the spectroscopic
survey).  In Section \ref{results} we discuss the impact of this
selection on the results of the paper.

\subsection{Atmospheric parameters}\label{atmpar}

\begin{figure}%[!htb]
\includegraphics[height=12cm,width=15cm]{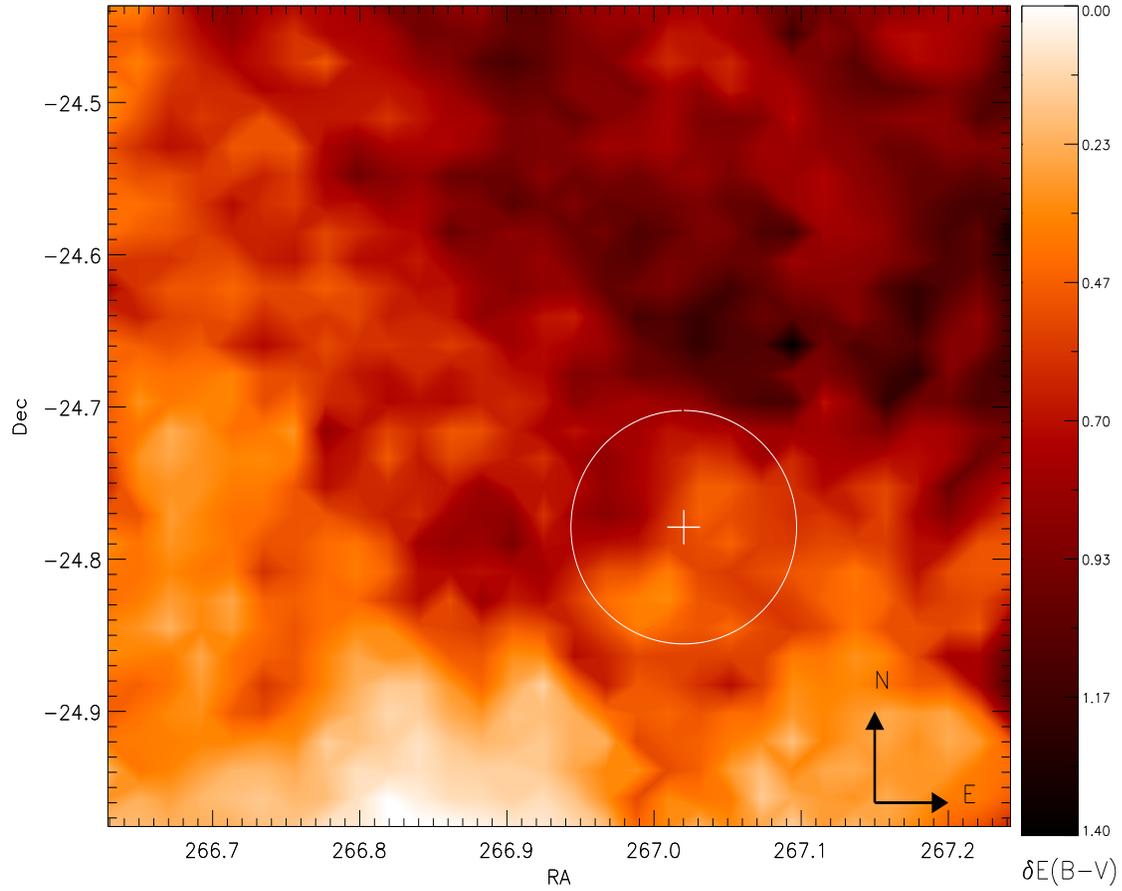}
\caption{\small Reddening map in the direction of Terzan 5 covering
  the entire $\sim 25$\arcmin$\times 25$\arcmin ~FoV.  Dark colors
  correspond to regions of large extinction (see the
  color bar on the right).  The center and tidal radius of Terzan 5
  are marked for sake of comparison.}
\label{redmap}
\end{figure}

We derived effective temperatures (T$_{{\rm eff}}$) and gravities ($\log$ $g$) photometrically. 
In order to minimize the effect of differential reddening we used the 2MASS catalog, correcting the (K, J-K) CMD
for differential extinction according to our new wide-field reddening map, shown in Figure \ref{redmap}.
This was obtained by applying to the optical WFI catalog the same procedure described in \cite{massari}\footnote{A webtool to compute the reddening
in the direction of Terzan 5 is publically available at the cosmic-lab website, http://www.cosmic-lab.eu/Cosmic-Lab/Products.html}. 
Because of the large incompleteness at the Main Sequence (MS) level we were forced to use red clump stars as reference.
Since these stars are significantly less numerous than MS stars, the spatial resolution of the computed reddening map ($60\arcsec \times 60\arcsec$)
is coarser than that ($8\arcsec \times 8\arcsec$)  published in \cite{massari} for the HST ACS field of view. 
However, despite this difference in resolution, the WFI reddening map agrees quite well with that for ACS in the overlapping region.
Indeed for the stars in common between the two catalogs, the average difference between the differential reddening 
estimates is $\langle\Delta E(B-V)\rangle=0.01$ mag 
with a dispersion $\sigma=0.1$ mag. This latter  value is also the uncertainty that
we conservatively adopt for our color excess estimates.

Figure \ref{cmd} shows the IR color-magnitude diagram (CMD) after the internal reddening
correction, with the positions of the spectroscopic targets highlighted. 
To determine T$_{{\rm eff}}$, we adopted the (J-K)$_{0}-$T$_{{\rm eff}}$ empirical relation quoted by \cite{m98}.
Since the relation is calibrated onto the SAAO photometric system, we previously converted our 2MASS magnitudes
following the prescriptions in \cite{carpenter01}.
To estimate photometric gravities, we used the relation:
\begin{equation}
 \log g=\log g_\odot +4\log(T_{{\rm eff}}/T_{\odot})+\log(M/M_{\odot})+0.4(M_{{\rm bol}}-M_{{\rm bol},\odot})
\end{equation}
adopting $\log$ $g_{\odot}$=4.44 dex, T$_{\odot}$=$5770$ K, M$_{{\rm bol},\odot}=4.75$, M$=0.8$ M$_\odot$ and a distance of 8 kpc.
Such a distance is the average value predicted by the Besan\c{c}on model for a simulated field with the size of the FoV covered 
by our observations and centered around Terzan 5.  This value is also normally adopted when bulge stars are analyzed
(\citealt{zoccali08, alvesbrito10, hill11}, see also Sect. \ref{results} for a discussion on the impact of distance on our results).
Bolometric corrections were taken from \cite{m98}.
The small number (about 10) of Fe~I lines available in the spectra prevents us to derive reliable values of
microturbulent velocity ($v_{turb}$; see \citealt{m11} for a review of the different methods to infer this parameter).
We therefore referred to the works of \cite{zoccali08} and \cite{johnson13} on large samples of bulge giant stars 
characterized by metallicities and atmospheric parameters similar to those of our targets.
Since no specific trend between $v_{turb}$ and the atmospheric parameters is found in these samples, we adopted their 
median velocity $v_{turb}=$1.5 km$\,$s$^{-1}$ ($\sigma=0.16$ km$\,$s$^{-1}$) for all the targets.

\subsection{Chemical analysis}\label{analysis}

The Fe~I lines used for the chemical analysis were selected from the latest version of the Kurucz/Castelli dataset of atomic 
data\footnote{http://wwwuser.oat.ts.astro.it}. 
We included only Fe I transitions found to be unblended in synthetic spectra calculated 
with the typical atmospheric parameters of our targets and 
at the resolutions provided by the GIRAFFE and DEIMOS spectrographs. 
These synthetic spectra were calculated with the SYNTHE code, 
including the entire Kurucz/Castelli line-list (both for atomic and molecular lines) convolved with a 
Gaussian profile at the resolution of the observed spectra. 
Due to the different spectral resolution of the two datasets, we used two different techniques 
to analyze the spectral lines and determine the chemical abundances.

(1)~{\it FLAMES spectra}---
The chemical analysis was performed using the package GALA \citep{gala}\footnote{GALA 
is freely distributed at the Cosmic-Lab project website, 
http://www.cosmic-lab.eu/gala/gala.php}, an automatic tool to derive the chemical abundances 
of single, unblended lines by using their measured equivalent widths (EWs). 
The adopted model atmospheres were calculated with the ATLAS9 code \citep{atlas}.
In our analysis, we run GALA fixing all the atmospheric parameters estimated as described above and 
leaving only the metallicity of the model atmosphere free to vary iteratively in order to match the iron abundance
derived from EWs.
EWs were obtained with the code 4DAO (\citealt{4dao})\footnote{Also this code is freely distributed at the 
Cosmic-lab website: http://www.cosmic-lab.eu/4dao/4dao.php.}, 
aimed at running DAOSPEC \citep{daospec} for a large set of spectra, tuning automatically 
the main input parameters used by DAOSPEC and providing graphical outputs to visually 
inspect the quality of the fit for each individual spectral line.
The EWs were measured adopting a Gaussian function that is a reliable approximation for the line profile 
at the resolution of our spectra. EW errors were estimated by DAOSPEC 
from the standard deviation of the local flux residuals \citep[see][]{daospec} and lines with EW errors 
larger than 10\% were rejected.

(2)~{\it DEIMOS spectra}---
Due to the low resolution of the DEIMOS spectra,  
the high degree of line blending and blanketing makes the derivation of the abundances through the method of 
the EWs  quite complex and uncertain. Thus, the iron abundances were measured by 
comparing the observed spectra with a grid of synthetic spectra, 
following the procedure described in \citet{m12_2419}. Each Fe~I line was 
analyzed independently by performing a $\chi^2$-minimization between the normalized observed spectrum and 
a grid of synthetic spectra (computed with the code SYNTHE, convolved at DEIMOS resolution and 
resampled at the pixel-size of the observed spectra). Then, the normalization is readjusted locally 
in a region of $\sim$50-60 \AA{} in order to improve the quality of the fit\footnote{No systematic differences
in the iron abundances obtained from FLAMES and DEIMOS spectra have been found for the targets in common between the two datasets}. 
Uncertainties in the fitting procedure for each spectral line were estimated by using 
Monte Carlo simulations: for each line, Poissonian noise was added to the best-fit synthetic spectrum
in order to match the observed SNR, and then the fit was re-computed with the same procedure described 
above. A total of 1000 Monte Carlo realizations has computed for each line, and
the dispersion of the derived abundance distribution was adopted as the abundance uncertainty. 
Typical values are of about $\pm$0.2 dex. 

\subsection{Calibration stars}

Because of its prominence, the Ca~II triplet is commonly used as a
proxy of the metallicity. However, several Fe~I lines fall in the
spectral range of the adopted FLAMES and DEIMOS setups and we
therefore decided to measure the iron abundance directly from these
lines. To demonstrate the full reliability of the atomic data adopted
to derive the metallicity we performed the same analysis on a set of
high-resolution, high-SNR spectra of the Sun and of Arcturus.  For the
Sun we adopted the solar flux spectrum quoted by \citet{neckel}, while
for Arcturus we used the high-resolution spectrum of \cite{hinkle}.
Both spectra were analyzed by adopting the same linelist used for the
targets of this study. For the Sun, the solar model atmosphere
computed by
F. Castelli\footnote{http://wwwuser.oat.ts.astro.it/castelli/sun/ap00t5777g44377k1asp.dat}
was used (T$_{{\rm eff}}$=5777 K, $\log$~$g$=4.44 dex), and v$_{{\rm
    turb}}$=1.2 km$\,$s$^{-1}$ (\citealt{andersen99}) was adopted. For
Arcturus we calculated a suitable ATLAS9 model atmosphere with the
atmospheric parameters (T$_{{\rm eff}}$=5286 K, $\log$~$g$=1.66 dex,
v$_{{\rm turb}}$=1.7 km$\,$s$^{-1}$) listed by \citet{ramirez11}.  The
resulting iron abundance for the Sun is A(Fe~I)$_{Sun}$=7.49$\pm$0.03
dex, in very good agreement with that listed by
\citet[][A(Fe)$=7.50$]{gs98}.  For Arcturus we obtained
A(Fe~I)$_{Arcturus}$=7.00$\pm$0.02, corresponding to
[Fe/H]$=-0.50\pm0.02$ dex, in excellent agreement with the measure of
\citet{ramirez11} who quote [Fe/H]$=-0.52\pm0.02$.  Thus, we conclude
that the adopted atomic lines provide a reliable estimate of the iron
abundance.

As discussed in Section \ref{iron}, for the 158 stars with $0.6<q<0.8$
we limited the metallicity analysis to the iron lines in a restricted
wavelength range ($8680-8850$ \AA) poorly affected by the TiO bands.
In order to properly check for any possible systematic effect due to
the different line list adopted, we re-performed the metallicity
analysis of the 126 stars with $q>0.8$ using only the reduced line
list. We found a very small off-set in the derived abundance
($\delta[Fe/H]=$[Fe/H]$_{full}$-[Fe/H]$_{reduced}=-0.06\pm0.01$ dex), that
was finally applied to the iron abundance obtained for the 158 low-$q$
targets.

\subsection{Uncertainties}

The global uncertainty of our iron abundance estimates (typically
$\sim0.2$ dex) is computed as the sum in quadrature of two different
sources of error. The first one is the error arising from the
uncertainties on the atmospheric parameters. Since they have been
derived from photometry, the formal uncertainty on these quantities
depends on all those parameters which can affect the location of the
targets in the CMD, such as photometric errors, uncertainty on the
absolute and differential reddening and errors on the distance modulus
(DM). In order to evaluate the uncertainties on $T_{eff}$ and
$\log$~$g$ we therefore repeated their estimates for every single
target assuming $\sigma_{K}=0.04$, $\sigma_{J-K}=0.05$,
$\sigma_{\delta[E(B-V)]}=0.1$ (see Section \ref{atmpar}),
$\sigma_{[E(B-V)]}=0.05$ (\citealt{massari}) and a conservative value
$\sigma_{\rm DM}=0.3$ (corresponding to $\pm1$ kpc) for the DM.
Following this procedure we found uncertainties of $\pm160$ K in
T$_{{\rm eff}}$ and $\pm0.2$ dex in $\log$~$g$. For $v_{turb}$ we
adopted a conservative uncertainty of 0.2 km$\,$s$^{-1}$ (see Section
\ref{atmpar}).  To estimate the impact of these uncertainties on the
iron abundance, we repeated the chemical analysis assuming, each time, a
variation by 1$\sigma$ of any given parameter (keeping the other
ones fixed).

The second source of error comes from the internal abundance estimate
uncertainty. For each target this was estimated as the dispersion
around the mean of the abundances derived from the used lines, divided
by the root square of the number of lines. It is worth noticing that for
any given star the dispersion is calculated by weighting the abundance
of each line by its own uncertainty (as estimated by DAOSPEC for the
FLAMES targets, and from Monte Carlo simulations for the DEIMOS
targets).

\subsection{Results}
\label{results}
\begin{figure}%[!htb]
\plotone{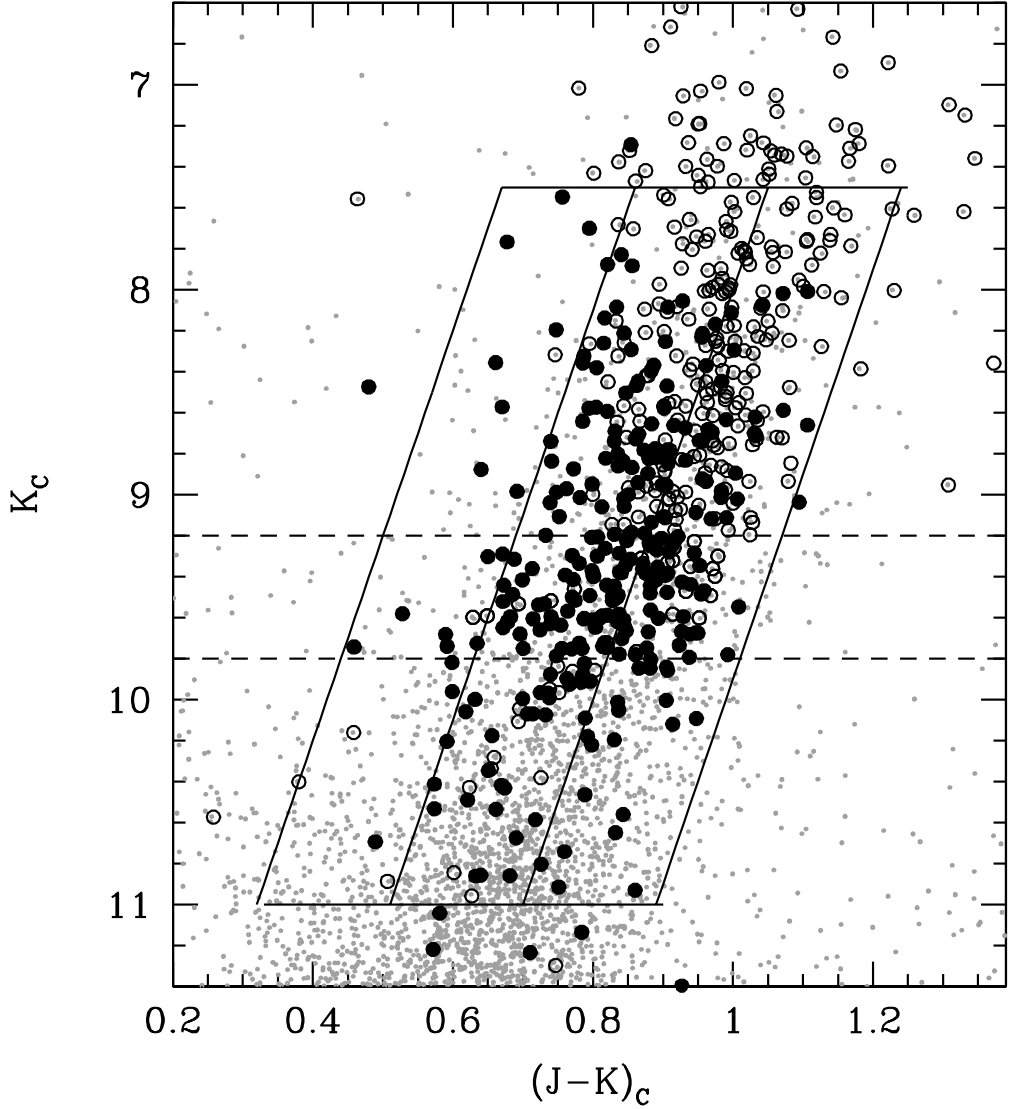}
\caption{\small(K, J-K) CMD corrected for internal differential
  extinction, for all the stars located at $400\arcsec<r<800\arcsec$
  from the center of Terzan 5, in our ESO-WFI photometric sample
  (small dots).  The spectroscopic targets are shown with large
  symbols. The targets for which iron abundance could be estimated are
  highlighted as large filled circles. The three strips adopted to
  evaluate the impact of the selection bias discussed in the text
  (Section \ref{results}) are shown. The horizontal dashed lines
  delimit the bias-free sample adopted to derive the metallicity
  distribution shown as a grey histogram  in Figure \ref{fedist}.}
\label{cmd}
\end{figure}

\begin{figure}%[!htb]
\plotone{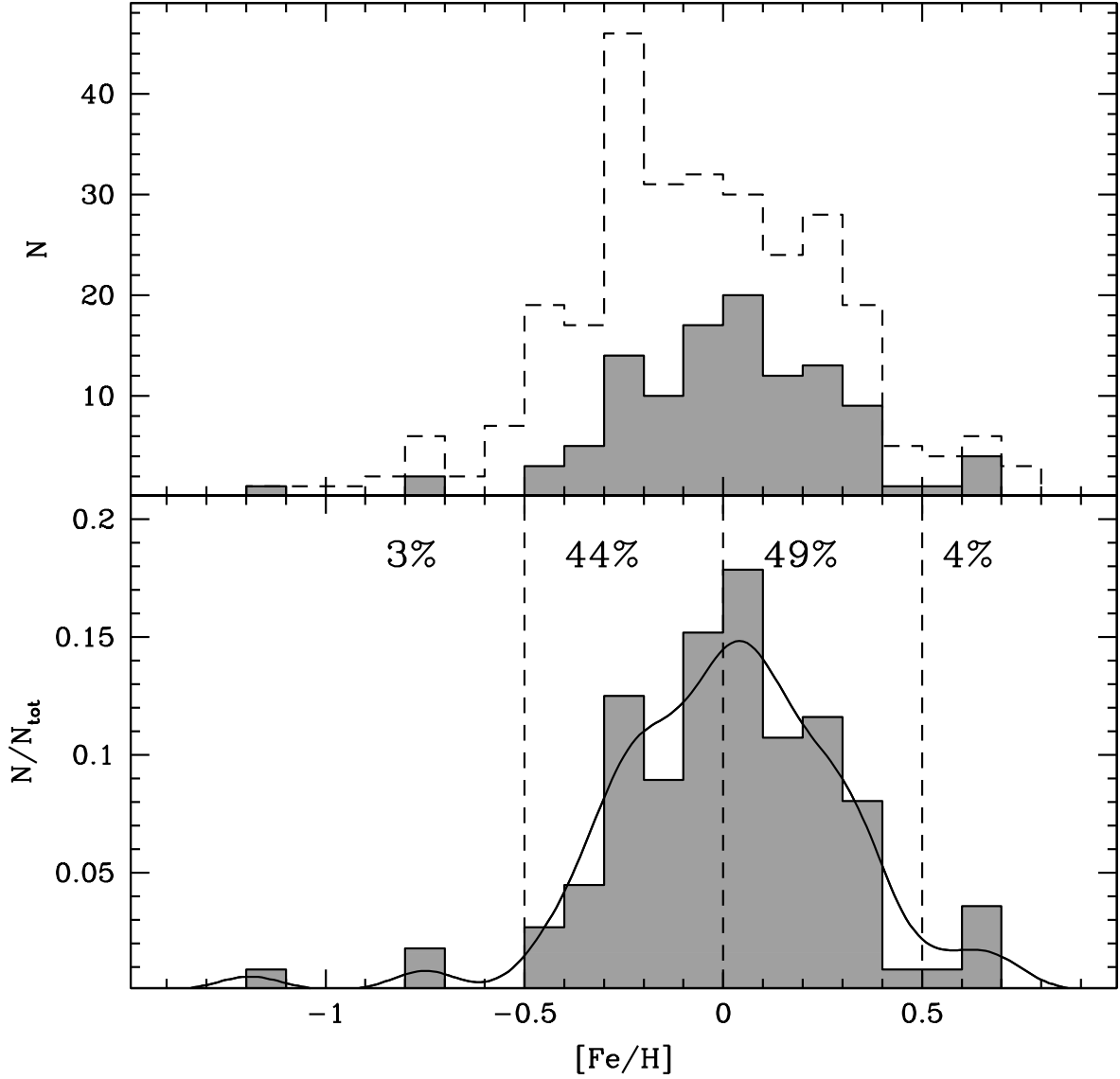}
\caption{\small {\it Top panel:} Metallicity distribution of the bulge
  field around Terzan 5 for the entire sample of 284 stars (dashed
  histogram) and for the sub-set of 112 targets selected at
  $9.2<K_c<9.8$ (grey histogram), free from the bias introduced by the
  TiO bands (which preferentially affects the spectra of the most
  metal-rich objects).  {\it Bottom panel:} Metallicity distribution
  observed in the unbiased sub-set of stars (the same grey histogram
  as above), compared to the generalized distribution (solid line)
  obtained from 1000 realization described in the text (Section
  \ref{results}).  The dashed lines delimit the metallicity ranges
  adopted to define the R$_{l/h}$ parameter (see text). The percentage
  of stars in each metallicity range is also marked.}
\label{fedist}
\end{figure}

The iron abundances and their total uncertainties for each of the 284
targets analyzed are listed in Table \ref{tab1}, together with the adopted
atmospheric parameters. The [Fe/H] distribution for the entire
sample is shown as a dashed-line histogram in Figure \ref{fedist}. The
distribution is quite broad, extending from [Fe/H]$\simeq-1.2$ dex, up
to [Fe/H]$\simeq0.8$ dex, with a pronounced peak at [Fe/H]$=-0.25$
dex. However, the exclusion of a significant fraction of stars with
spectra seriously contaminated by TiO bands (see Setc. \ref{iron}),
possibly introduced a selection bias on the derived metallicity
distribution.  In fact prominent TiO bands are preferentially expected
in the coolest and reddest stars.  This is indeed confirmed by Figure
\ref{cmd}, showing that the targets for which no abundance measure was
feasible (open circles) preferentially populate the brightest and
coolest portion of the RGB.  Since this is also the region were the
most metal-rich stars are expected to be found, in order to provide a
meaningful metallicity distribution, representative of the bulge
population around Terzan 5, we restricted our analysis to a sub-sample
of targets likely not affected by such a bias.  To this purpose, in
the CMD corrected for internal reddening we selected only stars in the
magnitude range $9.2<K_c<9.8$ (see dashed lines in Figure \ref{cmd}),
where metallicity measurements have been possible for 82\% of the
surveyed stars (i.e., 112 objects over a total of 136).  The
metallicity distribution for this sub-sample is shown as a grey
histogram in the top panel of Figure \ref{fedist}. The
distribution is still quite broad, extending from [Fe/H]$\simeq-1.2$
dex up to [Fe/H]$\simeq0.7$ dex, but the sub-solar component (with
$-0.5<$[Fe/H]$<0$ dex) seems to be comparable in size to the
super-solar component (with $0<$[Fe/H]$<0.5$ dex).
 
In order to properly evaluate the existence of any residual bias, we
followed the method described in \cite{zoccali08}.  We considered
three strips in the CMD roughly parallel to the slope of the bulge RGB
(Figure \ref{cmd}). In each strip, we computed the fraction $f$
defined as the ratio between the number of stars with measured
metallicity and the number of targets observed in the spectroscopic
surveys. In the selected sub-sample, the $f$ parameter ranges from
0.75 up to 0.90, with the peak in the central bin and with the reddest
bin being the less sampled. In order to evaluate the impact of this
residual inhomogeneity on the derived metallicity distribution we
randomly subtracted from the bluest and central bins a number of stars
(2 and 18, respectively) suitable to make the $f$ ratio constant in
all the strips. We repeated such a procedure $1000$ times, and for
each iteration a new metallicity distribution has been computed. The
bottom panel of Figure \ref{fedist} shows the generalized distribution
obtained from the entire procedure, overplotted to the observed one.
As can be seen, the two distributions are fully compatible, thus
providing the final confirmation that the observed distribution is not
affected by any substantial residual bias. Hence it has been adopted
as representative of the metallicity distribution of the bulge
population around Terzan 5.  

\begin{figure}%[!htb]
\plotone{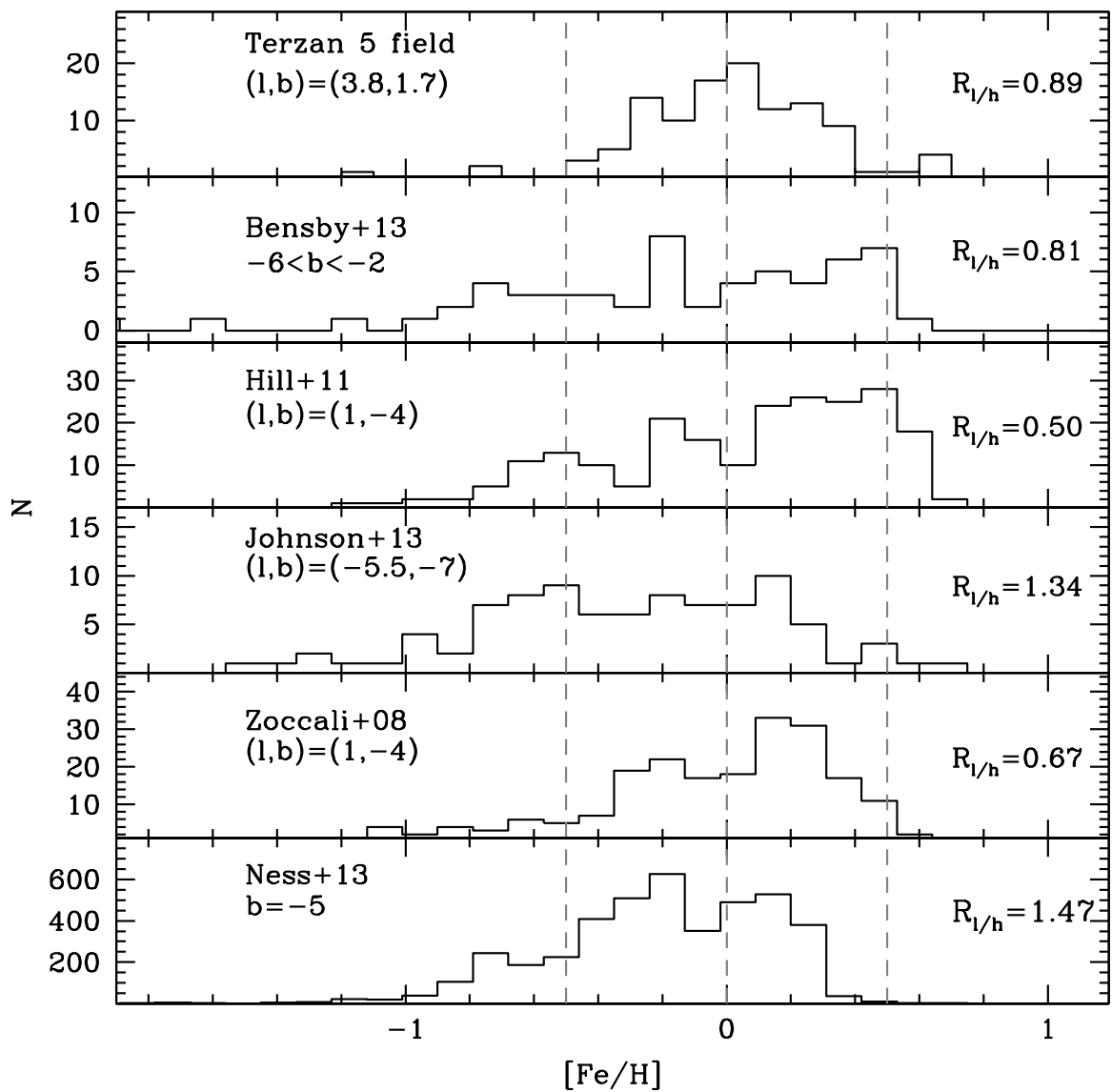}
\caption{\small Comparison of the iron distribution in a few bulge
  fields at different Galactocentric locations.  The corresponding
  references and Galactic coordinates are indicated in each
  panel. Vertical dashed lines delimitate the metallicity ranges
  defining the sub- and super-solar metallicity components.  The value
  of the R$_{l/h}$ parameter defined in the text (Section
  \ref{results}) is also reported in each panel. }
\label{distribs}
\end{figure}

In order to compare our results with previous studies, in Figure
\ref{distribs} we show the metallicity distribution obtained in the
present work and those derived in different regions of the Galactic
bulge: $-6$\textdegree$<$b$<-2$\textdegree \citep[for a sub-sample of
  micro-lensed dwarfs][]{bensby13}, the Baade's window (for a
sub-sample of giants; \citealp{hill11} and \citealp{zoccali08}),
l$=-5.5$\textdegree, b$=-7$\textdegree (for a sample of giants;
\citealp{johnson13}), and the \citet{ness13} field closest to the Galactic disk, at
b$=-5$\textdegree. The distributions appear quite
different. However a few common characteristics can be noted and
deserve a brief discussion. Apart from the presence of more or less
pronounced peaks, all the distributions show: ($i$) a major sub-solar
([Fe/H]$\simeq-0.2$ dex) component; ($ii$) a super-solar component
([Fe/H]$\simeq0.2$ dex); ($iii$) a quite extended tail towards low
metallicities (reaching [Fe/H]$\simeq-1.5$ dex). However, the relative
percentage in the two prominent components appears to be different
from one field to another. In order to properly quantify this feature,
we defined the ratio R$_{l/h}=$N$_{l}/$N$_{h}$, where N$_{l}$ is the
number of stars in the sub-solar component (with $-0.5<$[Fe/H]$<0$
dex) and N$_{h}$ is the number of stars in the super-solar component
(with $0<$[Fe/H]$<0.5$ dex). The value of R$_{l/h}$ is labelled in
each panel of Figure \ref{distribs}.

\begin{figure}%[!htb]
\plotone{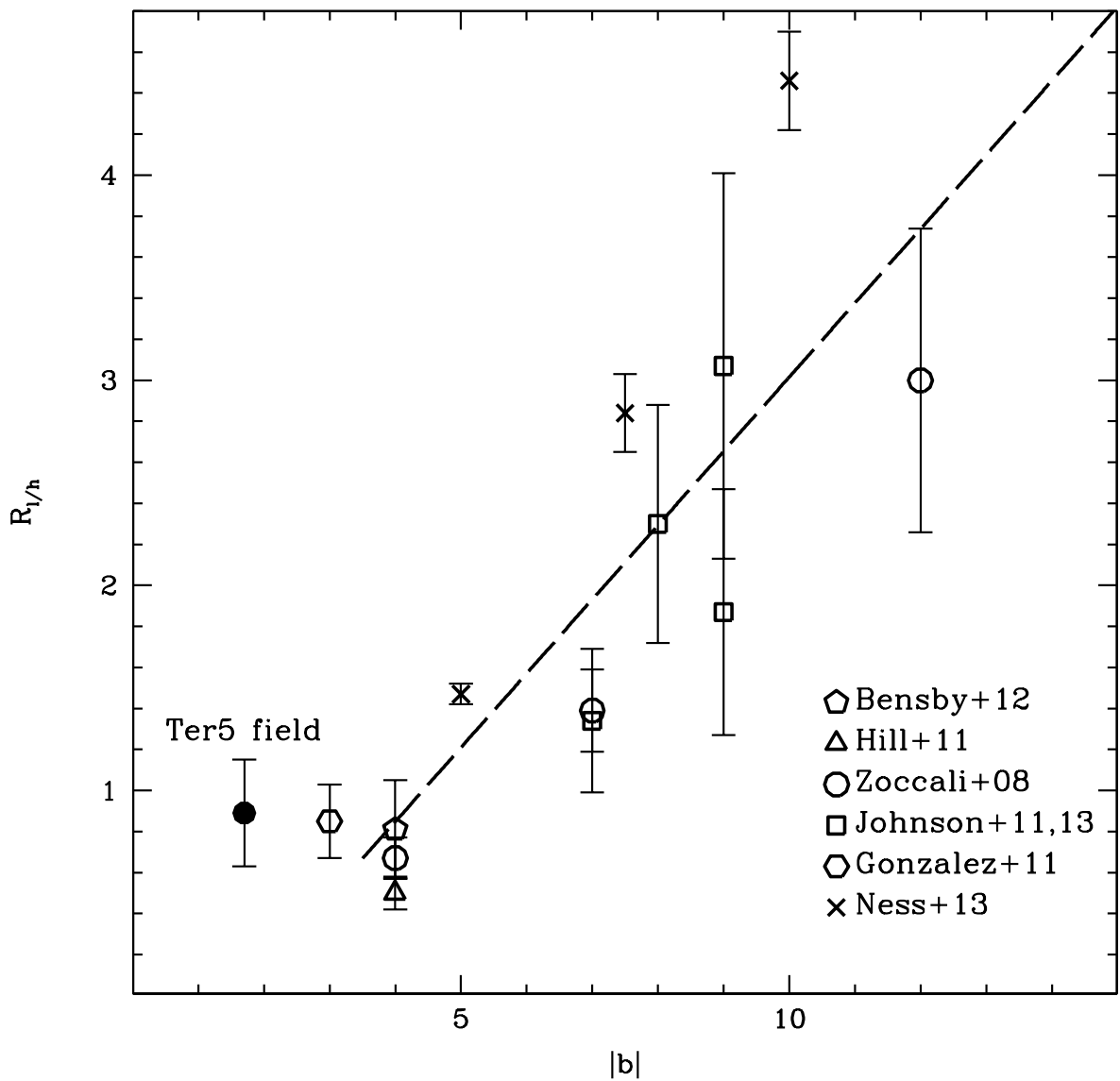}
\caption{\small R$_{l/h}$ parameter as a function of the Galactic
  latitude (absolute value).  The 13 fields taken from the literature
  nicely describe a metallicity gradient, suggesting that the
  super-solar component increases with decreasing Galactic
  latitude. The field measured around Terzan 5 is highlighted with a
  large filled circle and possibly suggests the presence of a
  ``plateau'' at R$_{l/h}\simeq 0.8$ for $|$b$|<4$\textdegree.  }
\label{grad}
\end{figure}

\begin{figure}%[!htb]
\plotone{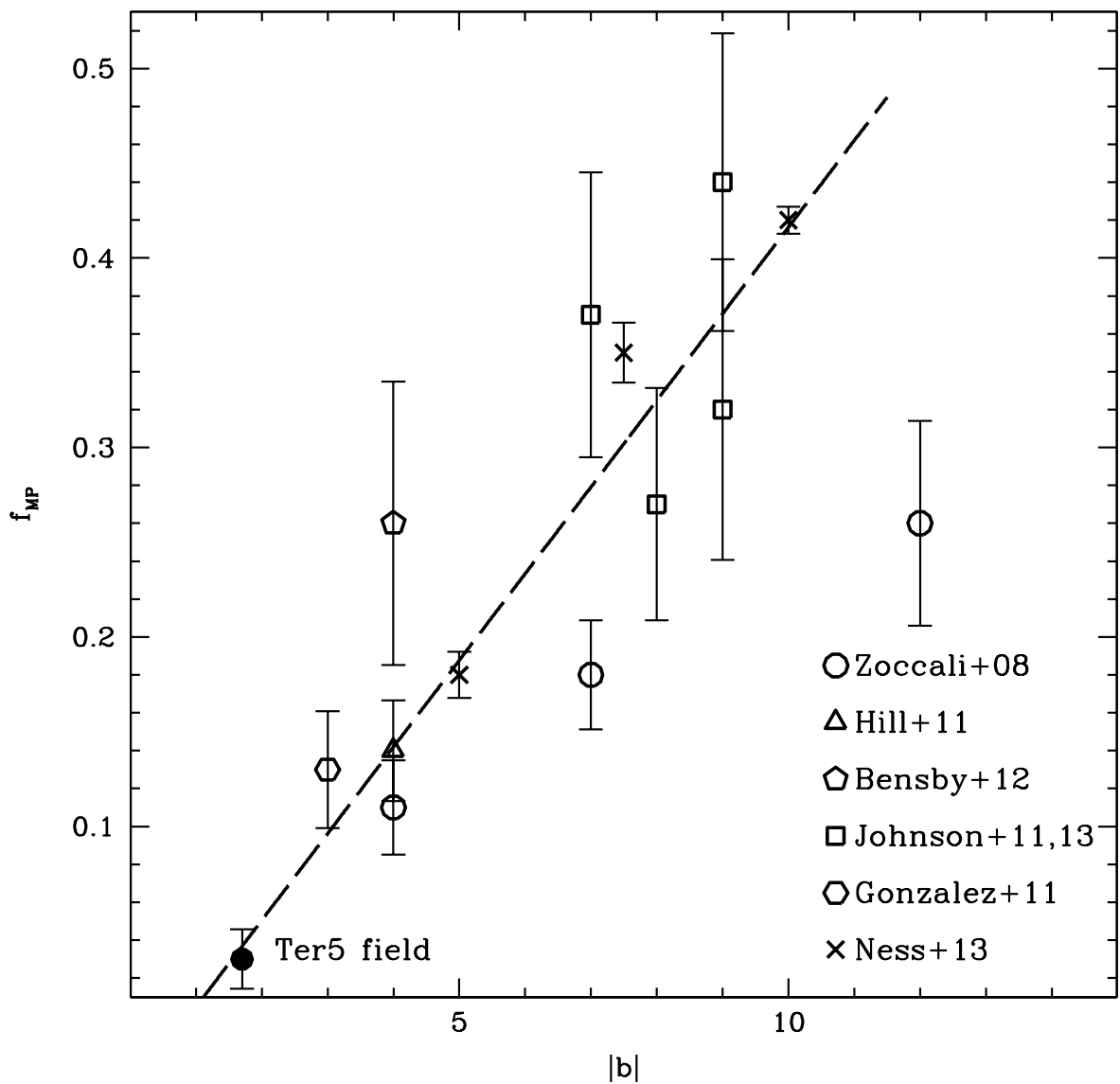}
\caption{\small Fraction of metal-poor stars with [Fe/H]$<-0.5$ dex (f$_{{\rm MP}}$) as a function of the
  absolute value of the Galactic latitude.  The considered bulge fields are the same as those
  in Fig. \ref{grad} and also in this case they describe a clear trend, with the only exception of
  the survey at the largest latitude (from \citealt{zoccali08}). The field measured around Terzan 5 
  is highlighted with a large filled circle and fits very well into the correlation.}
\label{gradpoor}
\end{figure}

In Figure \ref{grad} we plot the value of R$_{l/h}$ for 13 bulge
regions at different latitudes published in the literature (see
e.g. \citealt{bensby13}, \citealt{hill11}, \citealt{zoccali08},
\citealt{johnson11, johnson13}, \citealt{gonzalez11},
\citealt{ness13}). The value obtained for the bulge field around
Terzan 5 is shown as a large filled circle. It is interesting to note
how the populations observed in the 13 reference fields define a clear
trend, suggesting that the super-solar component tends to be dominant
at latitudes below $|$b$|<5$\textdegree.  The field around Terzan 5
is located at the lowest latitude observed so far. It
nicely fits into this trend, and it suggests the presence of a {\it
  plateau} at R$_{l/h}\sim 0.8$ for $|$b$|<4$\textdegree (see also
\citealt{rich12}). 
On the other hand, the Galactic location of the field can possibly be
the reason for the small amount of stars detected with [Fe/H]$<-0.5$.
Figure \ref{gradpoor} shows the fraction (f$_{{\rm MP}}$) of metal-poor
objects (with [Fe/H]$<-0.5$) with respect to the total number of stars
for each of the samples described in Figure \ref{grad}, as a function of
$|$b$|$.  Also in this case a clear trend is defined: the percentage of
metal-poor stars  drops from $\sim40$\% at $|$b$|\sim10$\textdegree~ to
a few percent at the latitude of Terzan 5, the only exception being
the most external field of \cite{zoccali08}.
Our findings are in good agreement with several recent results 
about the general properties of the Galactic bulge. Indeed, metal-rich 
stars are dominant at low Galactic latitudes, that is closer to the Galactic plane
\citep[see e.g.][and references therein]{ness14}.
Also, we checked the impact on these findings of the assumption of a $8$ kpc distance. 
We repeated the chemical analysis by adopting distances of $6$ and $10$ kpc. 
The change of the surface gravity (on average $+0.25$ dex and $-0.2$ dex, respectively) 
leads to only small differences in the measured 
[Fe/H], with the stellar metallicities differing on average by $0.06$ dex 
($\sigma=0.04$ dex) and $-0.05$ dex ($\sigma=0.02$ dex), in the two cases.
Such a tiny difference moves the value of R$_{l/h}$ from $0.89$ to $0.58$ and $1.26$, 
respectively, leaving it fully compatible with a flat behavior for 
$|$b$|<4$\textdegree~in both cases, while it does not change significantly 
(less than 1\%) the value of f$_{{\rm MP}}$.
We underline that this is the first determination of the 
{\it spectroscopic} metallicity distribution for a significant sample of stars 
at these low and positive Galactic latitudes.  Other spectroscopic surveys at 
low latitudes are needed to confirm the existence of these features.

Finally it is worth commenting on the velocity dispersion obtained for
the two main metallicity components in the field surrounding Terzan
5. We found two similar values, $\sigma_v=108\pm8$ km s$^{-1}$ and
$\sigma_v=111\pm11$ km s$^{-1}$ for the sub-solar and the super-solar
component, respectively.  The two measured values are in agreement
with those observed in the fields at b$=-5$\textdegree ~and at low
longitudes (l$<5$\textdegree) by \citet[][see the red diamonds in the
  lower panels of their Figure 7]{ness13kin} in the same metallicity
range (-0.5$<$[Fe/H]$<0.5$ dex).

\section{SUMMARY}
\label{summary} 
We determined the radial velocity distribution for a sample of 615
stars at (l,b)=(3.7\textdegree,1.8\textdegree), representative of the
bulge field population surrounding the peculiar stellar system Terzan
5. We found that the distribution is well fitted by a Gaussian
function with $<v_{\rm rad}>=21.0\pm4.6$ km s$^{-1}$ and
$\sigma_{v}=113.0\pm2.7$ km s$^{-1}$. Once converted to Galactocentric
velocities, these values are in agreement with the determinations
obtained in other bulge fields previously investigated.  We did
not find evidence for the high-velocity sub-component recently
identified in \cite{nidever}.

Because of the strong contamination of TiO bands, we were able to
measure the iron abundance only for a sample of 284 stars
(corresponding to $\sim 46\%$ of the entire sample) and we could
derive an unbiased metallicity distribution only from a sub-sample
of 112 stars with $9.2<K_c<9.8$. 
Statistical checks have been used to demonstrate that
this is a bias-free sample representative of the bulge population
around Terzan 5.  The metallicity distribution turns out to be quite
broad with a peak at [Fe/H]$\simeq+0.05$ dex and 
it follows the general metallicity-latitude trend
found in previous studies, with the number of super-solar bulge
stars systematically increasing with respect to the number of sub-solar ones
for decreasing latitude. Indeed the population ratio between the sub-solar
and super solar components (quantified here by the parameter R$_{l/h}$)
measured around Terzan 5 nicely agrees with that observed in other low
latitude bulge fields, possibly suggesting the presence of a plateau
for $|$b$|<4$\textdegree. Moreover, also the fraction of stars with
[Fe/H]$<-0.5$ measured around Terzan 5 fits well into the correlation 
with $|$b$|$ found from previous studies.
 
\acknowledgements{We thank the anonymous referee for his/her comments and 
  suggestions which helped us to improve the presentation
  of our results. This research is part of the project COSMIC-LAB
  (web site: http://www.cosmic-lab.eu) funded by the European Research
  Council (under contract ERC-2010-AdG-267675). 
  Some of the data presented herein were obtained at the W.M. Keck Observatory, 
  which is operated as a scientific partnership among the California Institute 
  of Technology, the University of California and the National Aeronautics and 
  Space Administration. The Observatory was made possible by the generous 
  financial support of the W.M. Keck Foundation.  R. Michael Rich acknowledges
  support from grant AST-1212095 from the US National Science Foundation.
  We warmly thank Melissa Ness for providing the metallicity measurements of the
  fields at b$=-5$\textdegree,$-7.5$\textdegree,$-10$\textdegree
  ~published in \cite{ness13}.  }

\begin{deluxetable}{rccccccc}
\tablewidth{0pc}
\tablecolumns{8}
\tiny
%\tablewidth{0pt}
\tablecaption{Identification number, coordinates, atmospheric parameters, iron abundances and
their uncertainties for the Terzan 5 field stars in our sample.}
\tablehead{\colhead{ID} & \colhead{RA} & \colhead{Dec} & \colhead{T$_{{\rm eff}}$} & \colhead{log~$g$} & \colhead{[Fe/H]} 
& \colhead{$\sigma_{[Fe/H]}$} & Dataset \\
& & & \colhead{(K)} & \colhead{(dex)} & \colhead{(dex)} & \colhead{(dex)} & }
\startdata
%\hline
 & & \\
%\hline
  1030711  &  267.1829348  &  -24.6923402  &  3922   &   0.6   &  -0.22    &   0.22   &  FLAMES \\
  1052484  &  267.1470436  &  -24.6830821  &  4220   &   0.8   &  -0.19    &   0.19   &  FLAMES \\
  1071029  &  267.1182806  &  -24.6825530  &  3971   &   1.9   &   0.39    &   0.18   &  FLAMES \\
  1071950  &  267.1169145  &  -24.6630493  &  3832   &   0.9   &   0.24    &   0.19   &  FLAMES \\
  1072160  &  267.1165783  &  -24.6588856  &  4111   &   0.9   &   0.63    &   0.20   &  FLAMES \\
  2009060  &  267.0686838  &  -24.6638585  &  4434   &   0.9   &  -0.34    &   0.17   &  FLAMES \\
  2029939  &  267.0366131  &  -24.6637204  &  4366   &   1.4   &   0.69    &   0.15   &  FLAMES \\
  2065353  &  266.9816844  &  -24.6414183  &  4013   &   0.9   &  -0.04    &   0.26   &  FLAMES \\
  2066891  &  266.9791962  &  -24.6560448  &  4371   &   1.1   &   0.22    &   0.19   &  FLAMES \\
  2068105  &  266.9771994  &  -24.6119254  &  3888   &   0.9   &   0.04    &   0.22   &  FLAMES \\
\enddata
\tablecomments{\small The entire table is available in the online version of the journal.}
\label{tab1}
\end{deluxetable}

\end{document}